\documentclass[10pt,twocolumn,aps,prb,superscriptaddress]{revtex4-1}
\usepackage{amsmath}
\usepackage{amssymb}
\usepackage{graphicx}

\usepackage{xspace}

\newcommand{\dvec}[1]{\ensuremath{\boldsymbol{#1}}\xspace}
\newcommand{\vk}{\dvec{\mathrm{k}}}
\newcommand{\vn}{\dvec{\mathrm{n}}}
\newcommand{\vq}{\dvec{\mathrm{q}}}

\newcommand{\vsigma}{\dvec{\sigma}}

\newcommand{\mos}{MoS$_2$\xspace}
\newcommand{\ws}{WS$_2$\xspace}

\newcommand{\eV}{\ensuremath{\mathrm{eV}}\xspace}
\newcommand{\cmsq}{\ensuremath{\mathrm{cm}^{-2}}\xspace}

\begin{document}

\author{K. Kechedzhi}
\affiliation{Condensed Matter Theory Center, 
	University of Maryland, College Park, MD 20742, USA}
\affiliation{Joint Quantum Institute, 
	University of Maryland, College Park, MD 20742, USA}

\author{D. S. L. Abergel}
\affiliation{Condensed Matter Theory Center, 
	University of Maryland, College Park, MD 20742, USA}
\affiliation{Nordita, 
	KTH Royal Institute of Technology and Stockholm University,
	Roslagstullsbacken 23, SE-106 91 Stockholm, Sweden}

\title{Weakly damped acoustic plasmon mode in transition metal
dichalcogenides with Zeeman splitting}

\begin{abstract}
We analyze the effect of a strong Zeeman field on the spectrum of
collective excitations of monolayer transition metal dichalcogenides.
The combination of the Dresselhaus type spin orbit coupling and an
external Zeeman field result in the lifting of the valley degeneracy
in the valence band of these crystals. We show that this lifting of
the valley degeneracy manifests in the appearance of an additional
plasmon mode with linear in wavenumber dispersion 
along with the standard square root in wavenumber
mode. Despite this novel mode being subject to the Landau damping,
it corresponds to a well defined quasiparticle peak in the spectral
function of the electron gas. 
\end{abstract}

\maketitle

A growing list of novel atomically thin crystals demonstrate great
potential for a wide variety of electronic applications due to the
exceptional tunability of their electronic properties and unique band
structure effects.\cite{geim_van_2013} In particular, monolayers
of a family of transition metal dichalcogenide (TMD) materials
of the form $\mathrm{AX}_{2}$ where $\mathrm{A=Mo,W,Ta}$ and
$\mathrm{X=S,Se,Te}$, have two-dimensional (2D) honeycomb lattice
structure and a well defined direct band gap $\gtrsim1.5\mathrm{eV}$
suitable for the use in transistors and other logical
elements. \cite{gordon_structures_2002, wang_electronics_2012,
radisavljevic_single-layer_2011}
Atomically thin crystals also show great promise for applications
in plasmonic devices. \cite{grigorenko_graphene_2012} 
This motivated a number of measurements of plasmon modes in
graphene \cite{liu_plasmon_2008, strait_confined_2013,
fei_gate-tuning_2012} and topological insulator thin films
\cite{pietro_observation_2013} and numerous device proposals. 
An extensive theory of the collective modes in 2D materials has been
reported \cite{hwang_dielectric_2007, hwang_plasmon_2009, 
sensarma_dynamic_2010, gangadharaiah_charge_2008, wunsch_dynamical_2006}
including the recent theoretical analysis of the spectrum of collective
modes of $\mathrm{MoS_{2}}$ \cite{scholz_plasmons_2013} which described
the typical 2D plasmon with a square root in wavenumber
dispersion in great detail. In this paper, we show that the collective
excitation spectrum of TMDs acquires a particularly rich structure
in the presence of Zeeman field, not included in the analysis in
Ref.~\onlinecite{scholz_plasmons_2013}.

One feature which distinguishes monolayer TMDs from other 2D hexagonal
crystals is that the partially filled $d$-orbitals of the heavy
transition metal atoms are characterized by a relatively strong
spin-orbit coupling (SOC). 
Also, the crystal lattice of a monolayer TMD breaks inversion symmetry
which results in the Dresselhaus-type~\cite{dresselhaus_spin-orbit_1955}
spin splitting which is particularly strong in the valence band
$\sim0.1-0.5\mathrm{eV}$. 
The SOC preserves the out-of-plane component of the spin as a good
quantum number. 
Low-energy electronic excitations in TMD materials are confined to
the close vicinity of the corners of the hexagonal Brillouin zone,
called K-points or valleys, and these valleys are related by the time
inversion operation.
Time inversion symmetry requires SOC to have opposite signs
in the two valleys, so that the spin up and down bands acquire energy
shifts of the opposite sign in the two valleys. In other words the
spin and valley indexes are locked at the top of the valence band,
resulting in valley Hall effect and valley dependent optical selection
rules. \cite{mak_control_2012, cao_valley-selective_2012,
zeng_valley_2012, xiao_coupled_2012}
This spin-valley locking also results in the long spin coherence time 
\cite{mak_control_2012, cao_valley-selective_2012, zeng_valley_2012}
which is expected to be limited only by the typically weak inter-valley
scattering on magnetic impurities or spin-lattice relaxation.
\cite{ochoa_spin_2013}

The contribution of the work presented here is to describe the rich
physics introduced by a non-zero Zeeman field, which results in a
relative energy shift of the spin polarized (spin-valley locked) bands
in the two valleys. 
The result is that when the Fermi level (which can be controlled by
external gating) is at the top of the valence band,
all the charge carriers (holes) are located in only one
valley which therefore demonstrates a chiral pseudospin texture similar
to the spin texture of a 3D topological insulator surface state with
a small gap. \cite{hasan_colloquium:_2010} In this case the plasmon
spectrum demonstrates a particularly rich evolution as a function
of the Fermi energy and/or Zeeman field strength. At higher hole doping
of the valence band the second valley becomes partially filled with
substantial carrier density imbalance between the two valleys. 
The density imbalance gives rise to a sizable difference of the Fermi
velocities in the two valleys. Coulomb interactions in the two bands
characterized by substantially different Fermi velocities give rise to a
novel plasmon mode with the frequency almost linear in wavenumber in
addition to the typical two dimensional plasmon mode with the square
root in wavenumber dispersion. 

An analogy can be drawn with a double layer semiconductor 2D electron
gas (2DEG) system considered by Das Sarma and Madhukar over 3 decades
ago.\cite{das_sarma_collective_1981} In this case Coulomb interactions
in the system of two 2DEGs physically separated by an insulator with
different electron densities give rise to a two-mode plasmon spectrum
with one mode having an almost linear in wavenumber dispersion and the
other the square root dispersion. \cite{das_sarma_collective_1981,
santoro_acoustic_1988, sensarma_dynamic_2010,
profumo_double-layer_2012}
The two modes correspond to symmetric and asymmetric combinations
of plasmons in each of the two layers. A crucial distinction with
TMDs considered here is that the two interacting 2DEGs in the case
of TMDs are located in the same physical space and differ only by
spin and valley quantum numbers. Furthermore, the presence of four
valence bands gives rise to a much richer plasmon spectrum in the
case of TMDs as compared to semiconductor double layer systems. 
Also, very recently a somewhat similar effect was
predicted~\cite{pisarra_acoustic_2013} in the case of a very highly
doped graphene, such that the Fermi level lies in the vicinity of the van
Hove singularity.
In the latter case, the strong anisotropy of the Fermi surface gives
rise to the large variation of the Fermi velocity and the resulting
splitting of the plasmon spectrum. 

In this paper, we consider a monolayer TMD in presence of a strong
Zeeman field.
We analyze the spectrum of collective excitations in the system in
the presence of Coulomb interaction within the random phase approximation
(RPA) of the many body perturbation theory. We analyze the splitting
of the plasmon spectrum in two or three modes caused by the Zeeman
field as a function of chemical potential. The novel linear modes
are subject to Landau damping due to electron-hole excitations. However,
we find that the damping is relatively weak which is reflected in
well defined non-Lorentzian peaks in the spectral function of the
electron gas associated with the novel plasmon modes. We also outline
the conditions for the existence of the linear modes.

The lattice symmetry of TMDs is similar to that of monolayer graphene
with the exception of the broken inversion symmetry which allows the
Dresselhaus type spin orbit coupling. Using the analogy with graphene
we can write the low energy $\vk \cdot \dvec{\mathrm{p}}$ Hamiltonian of a
TMD as
\begin{equation}
	H= \xi at \vk\cdot\vsigma + \frac{\Delta}{2}\sigma_{z}
	- \xi\lambda s_{z}\frac{\sigma_{z}-1}{2}+\alpha s_{z},
\end{equation}
where the momentum $k$ is defined with respect to K-points, $\xi=\pm1$
denotes the valley, and $\sigma_{i}$
and $s_{i}$ are Pauli matrices in the sublattice and spin spaces,
respectively. 
The first term is the nearest-neighbor hopping parametrized by the
lattice constant $a$ and the hopping energy $t$, the second term is the
band gap $\Delta$ introduced by the asymmetry between the two lattice
sites, and the third term is the spin-orbit coupling originating from
the $d$-orbitals of the metal ions parametrized by $\lambda$. Crucially,
we also allow a Zeeman term parametrized by $\alpha$. This Hamiltonian
can also be obtained by the expansion of the tight binding model
Hamiltonian~\cite{cappelluti_tight-binding_2013,
Rostami_Moghaddam_Asgari_2013, 
Kormanyos_Zolyomi_Drummond_Rakyta_Burkard_Falko_2013}
in the parameter $ka\ll1$. The energy spectrum consists of four bands
in each valley. The conduction bands are split only by the Zeeman
term, whereas the valence bands are split by a combination of the
Zeeman and spin-orbit terms: 
\begin{equation}
	E_{s\xi bk} = \frac{s\xi\lambda}{2} + s\alpha
	+ \frac{b}{2}\sqrt{4a^{2}t^{2}k^{2}+(\Delta-s\xi\lambda)^{2}},
\end{equation}
where $s=\pm1$ denotes the spin and $b=\pm1$ the conduction or valence
band.

\begin{figure}[tb]
	\centering 
	\includegraphics{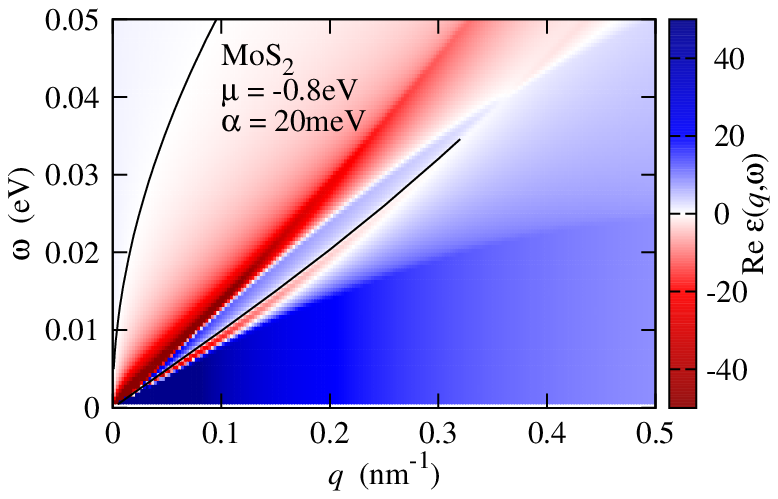}
	\includegraphics{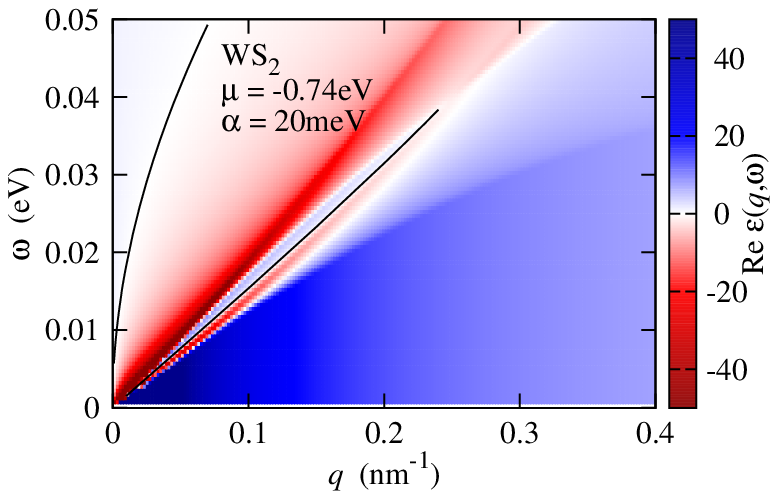}
	\caption{The real part of the dielectric function of (a) \mos and
	(b) \ws as a function of $q$ and $\omega$. The solid black lines
	mark the plasmon modes. 
	\label{fig:2dplots}}
\end{figure}

To compute the dielectric function and hence the plasmon spectrum, we
begin with the non-interacting polarization operator which is given by 
\begin{multline}
	\chi^{(0)}(\vq,\omega) = \int\frac{d^{2}\vk}{4\pi^{2}}
		\sum_{s,\xi,b} F_{s\xi b}(\vk,\vk+\vq) \times \\
	\times \frac{f_{s\xi bk}-f_{s\xi b|\vk+\vq|}}
		{\omega+i\eta+ E_{s\xi bk} - E_{s\xi b|\vk+\vq|}}.
	\label{eq:Chi0}
\end{multline}
In this expression, the functions $f_{s\xi bk}$ give the occupancy
of the state with the respective band indices and wave vector, and
$F_{s\xi b}(\vk,\vk+\vq)$ is the overlap of the wave functions in
the band denoted by $s$, $\xi$ and $b$ with wave vectors $\vk$
and $\vk+\vq$. Using the shorthand notation $\vk'=\vk+\vq$, this
is, 
\begin{multline}
F_{s\xi b}(\vk,\vk')=V_{s\xi bk}^{2}V_{s\xi bk'}^{2}+W_{s\xi
bk}^{2}W_{s\xi bk'}^{2}\\
+2V_{s\xi bk}V_{s\xi bk'}W_{s\xi bk}W_{s\xi bk'}\cos(\theta-\theta'),
\end{multline}
where $\theta$ and $\theta'$ are respectively the angle of the wave
vectors $\vk$ and $\vk'$ measured from the $k_{x}$ axis, and the
components of the spinor part of the wave function are 
\[
V_{s\xi bk}=\frac{-\xi
ak}{\sqrt{a^{2}t^{2}k^{2}+(\frac{\Delta}{2}+\alpha- E_{s\xi bk})^{2}}},
\]
and 
\[
	W_{s\xi bk}=\frac{\frac{\Delta}{2}+\alpha-E_{s\xi bk}}
		{\sqrt{a^{2}t^{2}k^{2}+(\frac{\Delta}{2}+\alpha-E_{s\xi bk})^{2}}}.
\]
The energy range that we are interested in is much smaller than the
band gap, and as such we can approximate the polarization operator
by the intra-band part originating from the two highest energy branches
of the valence band. These are the up spin band in the $K$ valley, and
the down spin band in the $K'$ valley. Note that the interband terms
in Eq.~(\ref{eq:Chi0}) arise only from the transitions between the
band with the same out-of-plane spin components due to orthogonality
of the opposite spin states. Therefore the largest interband term
is of the order $\omega/\Delta\ll1$.

The dielectric function within the RPA is then found from the
polarization operator in the standard way:
\begin{equation}
	\epsilon^{\mathrm{RPA}}(\vq,\omega) = 
	1 - V_{q}\chi^{(0)}(\vq,\omega),
	\label{eq:epsilon}
\end{equation}
with $V_{q}=2\pi e^{2}/\kappa q$ where $\kappa$ is the effective
dielectric constant of the medium in which the 2D crystal is embedded.

We focus on the situation where the Zeeman term is finite so that
the up-spin and down-spin bands in opposite valleys have asymmetric
Fermi surfaces. The numerical evaluation of Eq.~\eqref{eq:epsilon}
is shown in Fig.~\ref{fig:2dplots}, along with the associated plasmon
frequencies. We use material parameters corresponding to \mos and
\ws, with the latter exhibiting rather larger spin-orbit coupling
due to the increased weight of the tungsten ion over the molybdenum
ion. The Hamiltonian parameters for each of these materials are shown
in Tab.~\ref{tab:matpar}. The regular 2D plasmon with $\sqrt{q}$
dispersion is seen as the higher frequency mode, but we focus on the
emergence of a second mode at lower frequency with almost linear dispersion.
This mode is absent in the $\alpha=0$ case. Both modes have frequency
which increases with the hole density. 

\begin{table}[tb]
\begin{tabular}{|c|cccc|}
\hline 
Material  & $a$ (\AA)  & $t$ (\eV)  & $\Delta$ (\eV)  & $\lambda$ (\eV) \tabularnewline
\hline 
\mos  & $3.193$  & 1.10  & 1.66  & 0.075 \tabularnewline
\ws  & $3.197$  & 1.37  & 1.79  & 0.215 \tabularnewline
\hline 
\end{tabular}\caption{Material parameters for \mos and \ws, from 
Ref.~\onlinecite{xiao_coupled_2012}.
We assume the average dielectric constant to be $\kappa=5$ throughout
the text. \label{tab:matpar}}
\end{table}

The emergence of the linear in wavenumber plasmon mode can be understood
in more detail analyzing long wavelength asymptotic, $q/k_{F}\ll1$.
In this limit we can approximate the wave function overlap factor
as $F_{s\xi b}(\mathbf{k},\mathbf{k+q})\approx1+O(q/k_{F})$. In
the leading order in $q/k_{F}$ we write for the intra-band polarization
operator of Eq.~(\ref{eq:Chi0})
\begin{align}
	\chi^{(0)}(\vq,\omega) &= -\int\frac{d^{2}k}{(2\pi)^{2}}
		\sum_{s} \frac{ \vn_{k}\cdot \vq 
		\left| \frac{\partial \varepsilon_{sk}}{\partial k}\right|
		\delta (\mu - \varepsilon_{sk})}
		{\omega+\vn_k\cdot \vq
		\left|\frac{\partial\varepsilon_{sk}}{\partial k} \right|+i\delta} 
		\nonumber \\
	&= -\sum_{s} \nu_{s}\left[ 
		\mathcal{I} \left(\frac{\omega}{qv_{Fs}}\right) 
		+i \mathcal{J} \left(\frac{\omega}{qv_{Fs}}\right)\right],
		\label{eq:Chi0CosineII}
\end{align}
where $\vn_{k} = \vk/k$, and 
$v_{Fs} = \left|\frac{\partial\varepsilon_{sk}}{\partial
k}\right|_{k=k_F}$
and $\nu_{Fs} = \frac{k_{Fs}}{2\pi v_{Fs}}$ are the absolute value
of the dispersion slope and the density of states at the Fermi level
in the band labeled by the spin index $s$, respectively. Here we
include only the valence band $b=-1$ contribution, and take into
account the spin-valley locking $\xi=s=\pm1$ so that $\varepsilon_{sk} =
E_{ss-k}$. In Eq.~(\ref{eq:Chi0CosineII}) we introduced
\begin{align*}
	\mathcal{I}(a) &= \int_{0}^{2\pi} \frac{d\varphi}{2\pi}
		\frac{\cos\varphi}{a+\cos\varphi}\\
	&= \theta\left(1-a\right) 
		- \theta\left(a-1\right)\left(\frac{a}{\sqrt{a^{2}-1}}-1\right),
\end{align*}
where the integral is taken in the sense of the principal value, and
\begin{equation*}
	\mathcal{J}(a) = -\pi\int\frac{d\varphi}{2\pi}
		\delta( a + \cos\varphi ) \cos \varphi
	  =-\frac{a}{\sqrt{1-a^{2}}}\theta(1-a).
\end{equation*}
Using the leading order expansion Eq.~(\ref{eq:Chi0CosineII}) we
look for zeros of the real part of the dielectric function therefore
\begin{equation}
	1 + V_{q}\sum_{s}\nu_{s}\mathcal{I}
	\left(\frac{\omega}{qv_{Fs}}\right)=0.
	\label{eq:DielZero}
\end{equation}

\begin{figure}
	\includegraphics[]{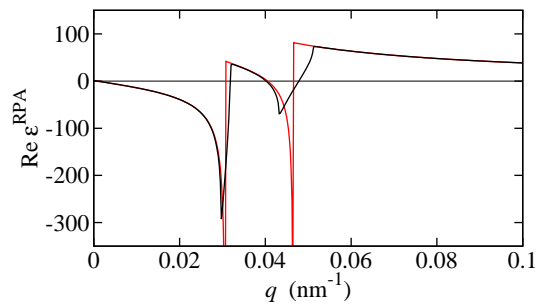}
	\caption{The real part of the dielectric function calculated using
	the full expression for the polarization in Eq.~(\ref{eq:epsilon}) 
	(black line) and the $q/k_F \ll 1$ approximation in
	Eq.~(\ref{eq:Chi0CosineII}) (red line).
	\label{fig:RealDiel2D}}
\end{figure}

The real part of the dielectric function at a fixed frequency is shown
in Fig.~\ref{fig:RealDiel2D} where the red line shows the approximation
by Eq.~(\ref{eq:Chi0CosineII}) and the black line is the full numerical
evaluation of Eq.~\eqref{eq:epsilon}. 
There is always a root of the dielectric function in the limit
$\frac{\omega}{v_{Fs}q}\gg1$ in which case Eq.\ (\ref{eq:DielZero})
reads
\begin{equation*}
	1-V_{q}\sum_{s}\frac{\nu_{s}\left(qv_{Fs}\right)^{2}}{2\omega^{2}}=0.
\end{equation*}
The solution corresponds to the plasmon mode with the dispersion given
by $\omega_{1}(q)=D\sqrt{q}$, where
\begin{align*}
	D &= \frac{e^{2}}{2\kappa}
		\left( k_{F\uparrow} v_{F\uparrow} 
		+ k_{F\downarrow}v_{F\downarrow}\right)\\
	& = \frac{e^{2}}{4\kappa} (2\mu -\lambda)
		\frac{2\left[ (2\mu -\lambda)^{2} - 4\alpha^{2}\right]
		- \left(\Delta-\lambda\right)^{2}}
		{\left(2\mu -\lambda\right)^{2}-4\alpha^{2}}.
\end{align*}

A second root of Eq.~(\ref{eq:DielZero}) may exist in the range
of frequencies $v_{F\downarrow}q\leq\omega\leq v_{F\uparrow}q$ given
by 
\[
	\omega_{2}(q) = v^{}_{F\downarrow} q 
	\frac{q+q_{\uparrow}+q_{\downarrow}}
	{\sqrt{\left(q+q_{\uparrow}+q_{\downarrow}\right)^{2} 
	- q_{\downarrow}^{2}}}\approx v_{F\downarrow}q,
\]
where $q_{s} = \frac{2\pi e^{2}}{\kappa}\nu_{s}$ is an analog
of Thomas-Fermi wavenumber for each of the spin/valley species. 
In order for the second root $\omega_{2}(q)$ to exist the dielectric
function has to become positive
\begin{equation}
	\varepsilon \left( \omega_{\uparrow},
	\frac{\omega_{\uparrow}}{v_{F\uparrow}}\right)>0,
	\label{eq:Diel2ndRottExistance}
\end{equation}
within $v_{F\downarrow}q\leq\omega\leq v_{F\uparrow}q$. The condition
in Eq.~(\ref{eq:Diel2ndRottExistance}) in the long wavelength limit,
$q/k_{F\downarrow}\ll1$, is satisfied so long as
\begin{equation*}
	q > q_{\downarrow}\frac{v_{F\uparrow} - \sqrt{v_{F\uparrow}^{2} 
	- v_{F\downarrow}^{2}}}{\sqrt{v_{F\uparrow}^{2}
	- v_{F\downarrow}^{2}}}-q_{\uparrow}.
\end{equation*}
In the lowest order in wavenumber $q/k_{F}\rightarrow0$ a divergence
appears in the dielectric function
\begin{equation}
	\lim_{a\rightarrow1^-}\mathcal{I}(a)\rightarrow-\infty,
	\label{eq:Divergence}
\end{equation}
which guarantees the existence of the second root. However,
at larger wavenumbers $q/k_{F\downarrow}\sim1$ the divergence in
Eq.~(\ref{eq:Divergence}) is rounded off and the second root of
the dielectric function disappears. This is reflected by the termination
of the linear branch of the plasmon dispersion shown in solid black
lines in Fig.~\ref{fig:2dplots}(a,b). At high hole doping the difference
in Fermi velocities becomes relatively small $v_{F\uparrow}\gg v_{F\uparrow}-v_{F\downarrow}$and
the linear mode disappears. 

\begin{figure}[tb]
\centering \includegraphics{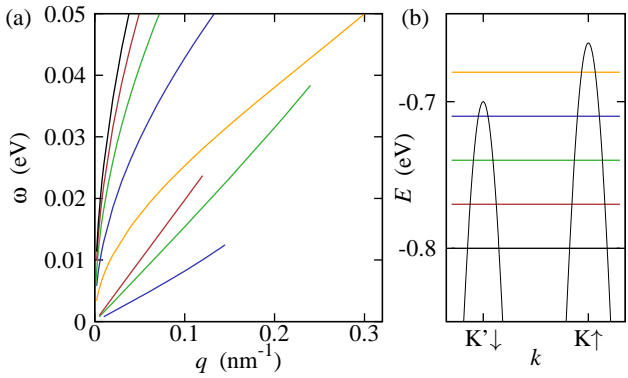}
\caption{(a) Variation of plasmon frequency with $\mu$ for \ws. (b) The two
partially occupied bands. The horizontal lines show the chemical potential
for each of the lines in (a). \label{fig:muvary}}
\end{figure}

The variation of the plasmon spectrum with the chemical potential
$\mu$ is shown in Figure \ref{fig:muvary}(a). When $\mu$ is located
in only the $\uparrow$-spin/$K$-valley band 
(orange line, $\mu=-0.68\eV$, $n=1.3\times10^{12}\cmsq$),
the second mode is absent. When the $\downarrow$-spin/$K'$-valley band 
becomes occupied (blue line, $\mu=-0.71\eV$, $n=2.0\times10^{12}\cmsq$)
the second mode becomes active. 
Initially, the range of momenta for which the second mode exists is
small and grows as $\mu$ moves deeper into the
$\downarrow$-spin/$K'$-valley band
(green line, $\mu=-0.74\eV$, $n=8.2\times10^{12}\cmsq$).
However, as the chemical potential increases further (red line,
$\mu=-0.77\eV$, $n=1.2\times10^{13}\cmsq$), the range of wave vectors
where the second mode exists starts to decrease, and eventually the
second mode vanishes (black line, $\mu=-0.8\eV$,
$n=1.7\times10^{13}\cmsq$).

\begin{figure}[tb] 
	\centering 
	\includegraphics{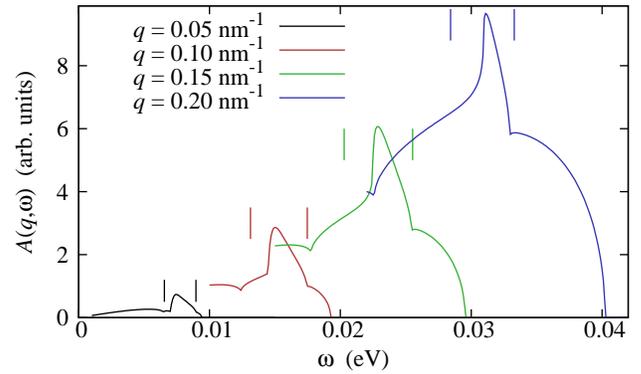}
	\caption{Spectral function of second plasmon mode is \ws. The
	vertical lines on either side of the peak denote the location of the
	two continua.  
	\label{fig:spectralfunction}} 
\end{figure}

The first plasmon mode $\omega_{1}(q)$ is not affected by Landau
damping, which is limited to the frequency range 
$\omega_{s}\leq v_{Fs}q$.
However, the second mode $v_{F\downarrow}q\leq\omega_{2}(q)\leq
v_{F\uparrow}q$ is subject to Landau damping due to electron-hole
excitations allowed in the $\uparrow$-spin/$K$-valley band. 
To illustrate the stability of the second plasmon mode, in
Fig.~\ref{fig:spectralfunction} we plot the spectral function
associated with the plasmon propagator 
\begin{equation*}
	A(\vq,\omega) = -\mathrm{Im}\chi^{\mathrm{RPA}}(\vq,\omega)
	= -\mathrm{Im}\frac{\chi^{(0)}(\vq,\omega)}
		{\epsilon^{\mathrm{RPA}}(\vq,\omega)},
\end{equation*}
as a function of $\omega$ for different $q$. The quasiparticle peak is
well defined at all wave vectors, and is located between the onset of
the two continua, in the narrow region of frequencies between the two
electron-hole excitation thresholds of $\uparrow$-spin and 
$\downarrow$-spin bands,
$v_{F\downarrow}q\leq\omega\leq v_{F\uparrow}q$, respectively. The line
shape of the quasiparticle peak corresponding to $\omega_{2}(q)$ is
distinct from a regular Lorentzian, which reflects Fano-like suppression
of the spectral weight due to mixing of plasmons with the electron-hole
excitation continuum. The strength of the response from the continuum
grows with $q$, but the height of the quasiparticle peak above the
continuum increases slightly. The width of the quasiparticle peak also
decreases with increasing $q$. Note that the ratio of the peak maximum
to the background of the spectral function is consistent with the real
and imaginary parts of the dynamical dielectric function being of the
same order, in which case this ratio is roughly equal to 2. This
indicates that the additional quasiparticle peak identified here is a
generic property of a 2D electron gas with lifted spin degeneracy.

With regard to the possibility to observe the predicted splitting
of the plasmon spectrum in TMDs experimentally, a crucial parameter
is the Zeeman energy. The simplest way to induce strong Zeeman splitting
is using magnetic field. The smallest Zeeman field considered in our
quantitative calculation, $4\mathrm{meV}$, corresponds to a very
high but experimentally accessible magnetic field of $B\lesssim20\mathrm{T}$
(assuming $g$-factor $\approx4$ for an out-of-plane field 
\cite{Rostami_Moghaddam_Asgari_2013}).
At such high fields Hall quantization is possible resulting in a discrete
spectrum in contrast to the continuous case considered here. In this
case the plasmon spectrum is modified due to mixing with cyclotron
modes. The splitting of the spectrum predicted here will persist and
will be observable in presence of Landau quantization. This situation
deserves a separate detailed study. 
Rotating the magnetic field so that it is in the plane of the TMD
removes the Landau quantization, but the anisotropy of the $g$-factor
means that a stronger magnetic field $B\approx 40\mathrm{T}$ is required
to achieve a Zeeman field of $4\mathrm{meV}$.
An alternative way to achieve very large Zeeman fields (without Hall
quantization) is using the proximity effect with a magnetic insulator.
Recently, a proximity induced Zeeman field was demonstrated in
$\mathrm{Bi_{2}Se_{3}/EuS}$ heterostructures with a Curie temperature
$\gtrsim10\mathrm{K}$. \cite{wei_exchange-coupling-induced_2013,yang_2013}
Alternative magnetic insulators can have much higher Curie temperatures
for example $500\mathrm{K}$ was reported recently for strained ultra-thin
($<10$ unit cells) $\mathrm{LaSrMnO}$ film on a $\mathrm{StTiO}$
substrate,\cite{boschker_high-temperature_2012} and $300\mathrm{K}$ of
a layered anti-ferromagnet $\mathrm{MnTe}$. \cite{Walther_1967}

In conclusion, we analyzed the evolution of the plasmon spectrum TMDs
in the presence of a strong Zeeman field as a function of the chemical
potential. We find that in a wide range of parameters there appears
a second plasmon mode characterized by almost linear in wavenumber
dispersion. This additional mode is subject to the effect of Landau
damping. However, the damping is relatively weak and the mode is associated
with a pronounced quasiparticle peak in the spectral function of the
electron gas. The shape of the plasmon peak in the spectral function
is distinctly non-Lorentzian and reflects the mixing of the mode with
electron-hole excitations. The predicted splitting in the collective
excitation spectrum is  a clear signature of a strong Zeeman field
and may be used to identify proximity effect in a TMD/ferromagnetic
insulator heterostructure.

\acknowledgments

We would like to thank Sankar Das Sarma and Euyheon Hwang for helpful
discussions. This work is supported by CMTC-LPS-NSA and US-ONR-MURI.

\bibliographystyle{apsrev4-1}
\bibliography{LitMoS2plasmons}

\begin{thebibliography}{33}%
\makeatletter
\providecommand \@ifxundefined [1]{%
 \@ifx{#1\undefined}
}%
\providecommand \@ifnum [1]{%
 \ifnum #1\expandafter \@firstoftwo
 \else \expandafter \@secondoftwo
 \fi
}%
\providecommand \@ifx [1]{%
 \ifx #1\expandafter \@firstoftwo
 \else \expandafter \@secondoftwo
 \fi
}%
\providecommand \natexlab [1]{#1}%
\providecommand \enquote  [1]{``#1''}%
\providecommand \bibnamefont  [1]{#1}%
\providecommand \bibfnamefont [1]{#1}%
\providecommand \citenamefont [1]{#1}%
\providecommand \href@noop [0]{\@secondoftwo}%
\providecommand \href [0]{\begingroup \@sanitize@url \@href}%
\providecommand \@href[1]{\@@startlink{#1}\@@href}%
\providecommand \@@href[1]{\endgroup#1\@@endlink}%
\providecommand \@sanitize@url [0]{\catcode `\\12\catcode `\$12\catcode
  `\&12\catcode `\#12\catcode `\^12\catcode `\_12\catcode `\%12\relax}%
\providecommand \@@startlink[1]{}%
\providecommand \@@endlink[0]{}%
\providecommand \url  [0]{\begingroup\@sanitize@url \@url }%
\providecommand \@url [1]{\endgroup\@href {#1}{\urlprefix }}%
\providecommand \urlprefix  [0]{URL }%
\providecommand \Eprint [0]{\href }%
\providecommand \doibase [0]{http://dx.doi.org/}%
\providecommand \selectlanguage [0]{\@gobble}%
\providecommand \bibinfo  [0]{\@secondoftwo}%
\providecommand \bibfield  [0]{\@secondoftwo}%
\providecommand \translation [1]{[#1]}%
\providecommand \BibitemOpen [0]{}%
\providecommand \bibitemStop [0]{}%
\providecommand \bibitemNoStop [0]{.\EOS\space}%
\providecommand \EOS [0]{\spacefactor3000\relax}%
\providecommand \BibitemShut  [1]{\csname bibitem#1\endcsname}%
\let\auto@bib@innerbib\@empty
\bibitem [{\citenamefont {Geim}\ and\ \citenamefont
  {Grigorieva}(2013)}]{geim_van_2013}%
  \BibitemOpen
  \bibfield  {author} {\bibinfo {author} {\bibfnamefont {A.~K.}\ \bibnamefont
  {Geim}}\ and\ \bibinfo {author} {\bibfnamefont {I.~V.}\ \bibnamefont
  {Grigorieva}},\ }\href {\doibase 10.1038/nature12385} {\bibfield  {journal}
  {\bibinfo  {journal} {Nature}\ }\textbf {\bibinfo {volume} {499}},\ \bibinfo
  {pages} {419} (\bibinfo {year} {2013})}\BibitemShut {NoStop}%
\bibitem [{\citenamefont {Gordon}\ \emph {et~al.}(2002)\citenamefont {Gordon},
  \citenamefont {Yang}, \citenamefont {Crozier}, \citenamefont {Jiang},\ and\
  \citenamefont {Frindt}}]{gordon_structures_2002}%
  \BibitemOpen
  \bibfield  {author} {\bibinfo {author} {\bibfnamefont {R.~A.}\ \bibnamefont
  {Gordon}}, \bibinfo {author} {\bibfnamefont {D.}~\bibnamefont {Yang}},
  \bibinfo {author} {\bibfnamefont {E.~D.}\ \bibnamefont {Crozier}}, \bibinfo
  {author} {\bibfnamefont {D.~T.}\ \bibnamefont {Jiang}}, \ and\ \bibinfo
  {author} {\bibfnamefont {R.~F.}\ \bibnamefont {Frindt}},\ }\href {\doibase
  10.1103/PhysRevB.65.125407} {\bibfield  {journal} {\bibinfo  {journal}
  {Physical Review B}\ }\textbf {\bibinfo {volume} {65}},\ \bibinfo {pages}
  {125407} (\bibinfo {year} {2002})}\BibitemShut {NoStop}%
\bibitem [{\citenamefont {Wang}\ \emph {et~al.}(2012)\citenamefont {Wang},
  \citenamefont {Kalantar-Zadeh}, \citenamefont {Kis}, \citenamefont
  {Coleman},\ and\ \citenamefont {Strano}}]{wang_electronics_2012}%
  \BibitemOpen
  \bibfield  {author} {\bibinfo {author} {\bibfnamefont {Q.~H.}\ \bibnamefont
  {Wang}}, \bibinfo {author} {\bibfnamefont {K.}~\bibnamefont
  {Kalantar-Zadeh}}, \bibinfo {author} {\bibfnamefont {A.}~\bibnamefont {Kis}},
  \bibinfo {author} {\bibfnamefont {J.~N.}\ \bibnamefont {Coleman}}, \ and\
  \bibinfo {author} {\bibfnamefont {M.~S.}\ \bibnamefont {Strano}},\ }\href
  {\doibase 10.1038/nnano.2012.193} {\bibfield  {journal} {\bibinfo  {journal}
  {Nature Nanotechnology}\ }\textbf {\bibinfo {volume} {7}},\ \bibinfo {pages}
  {699} (\bibinfo {year} {2012})}\BibitemShut {NoStop}%
\bibitem [{\citenamefont {Radisavljevic}\ \emph {et~al.}(2011)\citenamefont
  {Radisavljevic}, \citenamefont {Radenovic}, \citenamefont {Brivio},
  \citenamefont {Giacometti},\ and\ \citenamefont
  {Kis}}]{radisavljevic_single-layer_2011}%
  \BibitemOpen
  \bibfield  {author} {\bibinfo {author} {\bibfnamefont {B.}~\bibnamefont
  {Radisavljevic}}, \bibinfo {author} {\bibfnamefont {A.}~\bibnamefont
  {Radenovic}}, \bibinfo {author} {\bibfnamefont {J.}~\bibnamefont {Brivio}},
  \bibinfo {author} {\bibfnamefont {V.}~\bibnamefont {Giacometti}}, \ and\
  \bibinfo {author} {\bibfnamefont {A.}~\bibnamefont {Kis}},\ }\href {\doibase
  10.1038/nnano.2010.279} {\bibfield  {journal} {\bibinfo  {journal} {Nature
  Nanotechnology}\ }\textbf {\bibinfo {volume} {6}},\ \bibinfo {pages} {147}
  (\bibinfo {year} {2011})}\BibitemShut {NoStop}%
\bibitem [{\citenamefont {Grigorenko}\ \emph {et~al.}(2012)\citenamefont
  {Grigorenko}, \citenamefont {Polini},\ and\ \citenamefont
  {Novoselov}}]{grigorenko_graphene_2012}%
  \BibitemOpen
  \bibfield  {author} {\bibinfo {author} {\bibfnamefont {A.~N.}\ \bibnamefont
  {Grigorenko}}, \bibinfo {author} {\bibfnamefont {M.}~\bibnamefont {Polini}},
  \ and\ \bibinfo {author} {\bibfnamefont {K.~S.}\ \bibnamefont {Novoselov}},\
  }\href {\doibase 10.1038/nphoton.2012.262} {\bibfield  {journal} {\bibinfo
  {journal} {Nature Photonics}\ }\textbf {\bibinfo {volume} {6}},\ \bibinfo
  {pages} {749} (\bibinfo {year} {2012})}\BibitemShut {NoStop}%
\bibitem [{\citenamefont {Liu}\ \emph {et~al.}(2008)\citenamefont {Liu},
  \citenamefont {Willis}, \citenamefont {Emtsev},\ and\ \citenamefont
  {Seyller}}]{liu_plasmon_2008}%
  \BibitemOpen
  \bibfield  {author} {\bibinfo {author} {\bibfnamefont {Y.}~\bibnamefont
  {Liu}}, \bibinfo {author} {\bibfnamefont {R.~F.}\ \bibnamefont {Willis}},
  \bibinfo {author} {\bibfnamefont {K.~V.}\ \bibnamefont {Emtsev}}, \ and\
  \bibinfo {author} {\bibfnamefont {T.}~\bibnamefont {Seyller}},\ }\href
  {\doibase 10.1103/PhysRevB.78.201403} {\bibfield  {journal} {\bibinfo
  {journal} {Physical Review B}\ }\textbf {\bibinfo {volume} {78}},\ \bibinfo
  {pages} {201403} (\bibinfo {year} {2008})}\BibitemShut {NoStop}%
\bibitem [{\citenamefont {Strait}\ \emph {et~al.}(2013)\citenamefont {Strait},
  \citenamefont {Nene}, \citenamefont {Chan}, \citenamefont {Manolatou},
  \citenamefont {Kevek}, \citenamefont {Tiwari}, \citenamefont {{McEuen}},\
  and\ \citenamefont {Rana}}]{strait_confined_2013}%
  \BibitemOpen
  \bibfield  {author} {\bibinfo {author} {\bibfnamefont {J.~H.}\ \bibnamefont
  {Strait}}, \bibinfo {author} {\bibfnamefont {P.~S.}\ \bibnamefont {Nene}},
  \bibinfo {author} {\bibfnamefont {W.-M.}\ \bibnamefont {Chan}}, \bibinfo
  {author} {\bibfnamefont {C.}~\bibnamefont {Manolatou}}, \bibinfo {author}
  {\bibfnamefont {J.~W.}\ \bibnamefont {Kevek}}, \bibinfo {author}
  {\bibfnamefont {S.}~\bibnamefont {Tiwari}}, \bibinfo {author} {\bibfnamefont
  {P.~L.}\ \bibnamefont {{McEuen}}}, \ and\ \bibinfo {author} {\bibfnamefont
  {F.}~\bibnamefont {Rana}},\ }\href {http://arxiv.org/abs/1302.5972}
  {\bibfield  {journal} {\bibinfo  {journal} {{arXiv:1302.5972}}\ } (\bibinfo
  {year} {2013})}\BibitemShut {NoStop}%
\bibitem [{\citenamefont {Fei}\ \emph {et~al.}(2012)\citenamefont {Fei},
  \citenamefont {Rodin}, \citenamefont {Andreev}, \citenamefont {Bao},
  \citenamefont {{McLeod}}, \citenamefont {Wagner}, \citenamefont {Zhang},
  \citenamefont {Zhao}, \citenamefont {Thiemens}, \citenamefont {Dominguez},
  \citenamefont {Fogler}, \citenamefont {Neto}, \citenamefont {Lau},
  \citenamefont {Keilmann},\ and\ \citenamefont
  {Basov}}]{fei_gate-tuning_2012}%
  \BibitemOpen
  \bibfield  {author} {\bibinfo {author} {\bibfnamefont {Z.}~\bibnamefont
  {Fei}}, \bibinfo {author} {\bibfnamefont {A.~S.}\ \bibnamefont {Rodin}},
  \bibinfo {author} {\bibfnamefont {G.~O.}\ \bibnamefont {Andreev}}, \bibinfo
  {author} {\bibfnamefont {W.}~\bibnamefont {Bao}}, \bibinfo {author}
  {\bibfnamefont {A.~S.}\ \bibnamefont {{McLeod}}}, \bibinfo {author}
  {\bibfnamefont {M.}~\bibnamefont {Wagner}}, \bibinfo {author} {\bibfnamefont
  {L.~M.}\ \bibnamefont {Zhang}}, \bibinfo {author} {\bibfnamefont
  {Z.}~\bibnamefont {Zhao}}, \bibinfo {author} {\bibfnamefont {M.}~\bibnamefont
  {Thiemens}}, \bibinfo {author} {\bibfnamefont {G.}~\bibnamefont {Dominguez}},
  \bibinfo {author} {\bibfnamefont {M.~M.}\ \bibnamefont {Fogler}}, \bibinfo
  {author} {\bibfnamefont {A.~H.~C.}\ \bibnamefont {Neto}}, \bibinfo {author}
  {\bibfnamefont {C.~N.}\ \bibnamefont {Lau}}, \bibinfo {author} {\bibfnamefont
  {F.}~\bibnamefont {Keilmann}}, \ and\ \bibinfo {author} {\bibfnamefont
  {D.~N.}\ \bibnamefont {Basov}},\ }\href {\doibase 10.1038/nature11253}
  {\bibfield  {journal} {\bibinfo  {journal} {Nature}\ }\textbf {\bibinfo
  {volume} {487}},\ \bibinfo {pages} {82} (\bibinfo {year} {2012})}\BibitemShut
  {NoStop}%
\bibitem [{\citenamefont {Pietro}\ \emph {et~al.}(2013)\citenamefont {Pietro},
  \citenamefont {Ortolani}, \citenamefont {Limaj}, \citenamefont {Gaspare},
  \citenamefont {Giliberti}, \citenamefont {Giorgianni}, \citenamefont
  {Brahlek}, \citenamefont {Bansal}, \citenamefont {Koirala}, \citenamefont
  {Oh}, \citenamefont {Calvani},\ and\ \citenamefont
  {Lupi}}]{pietro_observation_2013}%
  \BibitemOpen
  \bibfield  {author} {\bibinfo {author} {\bibfnamefont {P.~D.}\ \bibnamefont
  {Pietro}}, \bibinfo {author} {\bibfnamefont {M.}~\bibnamefont {Ortolani}},
  \bibinfo {author} {\bibfnamefont {O.}~\bibnamefont {Limaj}}, \bibinfo
  {author} {\bibfnamefont {A.~D.}\ \bibnamefont {Gaspare}}, \bibinfo {author}
  {\bibfnamefont {V.}~\bibnamefont {Giliberti}}, \bibinfo {author}
  {\bibfnamefont {F.}~\bibnamefont {Giorgianni}}, \bibinfo {author}
  {\bibfnamefont {M.}~\bibnamefont {Brahlek}}, \bibinfo {author} {\bibfnamefont
  {N.}~\bibnamefont {Bansal}}, \bibinfo {author} {\bibfnamefont
  {N.}~\bibnamefont {Koirala}}, \bibinfo {author} {\bibfnamefont
  {S.}~\bibnamefont {Oh}}, \bibinfo {author} {\bibfnamefont {P.}~\bibnamefont
  {Calvani}}, \ and\ \bibinfo {author} {\bibfnamefont {S.}~\bibnamefont
  {Lupi}},\ }\href {\doibase 10.1038/nnano.2013.134} {\bibfield  {journal}
  {\bibinfo  {journal} {Nature Nanotechnology}\ }\textbf {\bibinfo {volume}
  {8}},\ \bibinfo {pages} {556} (\bibinfo {year} {2013})}\BibitemShut {NoStop}%
\bibitem [{\citenamefont {Hwang}\ and\ \citenamefont
  {Das~Sarma}(2007)}]{hwang_dielectric_2007}%
  \BibitemOpen
  \bibfield  {author} {\bibinfo {author} {\bibfnamefont {E.~H.}\ \bibnamefont
  {Hwang}}\ and\ \bibinfo {author} {\bibfnamefont {S.}~\bibnamefont
  {Das~Sarma}},\ }\href {\doibase 10.1103/PhysRevB.75.205418} {\bibfield
  {journal} {\bibinfo  {journal} {Physical Review B}\ }\textbf {\bibinfo
  {volume} {75}},\ \bibinfo {pages} {205418} (\bibinfo {year}
  {2007})}\BibitemShut {NoStop}%
\bibitem [{\citenamefont {Hwang}\ and\ \citenamefont
  {Das~Sarma}(2009)}]{hwang_plasmon_2009}%
  \BibitemOpen
  \bibfield  {author} {\bibinfo {author} {\bibfnamefont {E.~H.}\ \bibnamefont
  {Hwang}}\ and\ \bibinfo {author} {\bibfnamefont {S.}~\bibnamefont
  {Das~Sarma}},\ }\href {\doibase 10.1103/PhysRevB.80.205405} {\bibfield
  {journal} {\bibinfo  {journal} {Physical Review B}\ }\textbf {\bibinfo
  {volume} {80}},\ \bibinfo {pages} {205405} (\bibinfo {year}
  {2009})}\BibitemShut {NoStop}%
\bibitem [{\citenamefont {Sensarma}\ \emph {et~al.}(2010)\citenamefont
  {Sensarma}, \citenamefont {Hwang},\ and\ \citenamefont
  {Das~Sarma}}]{sensarma_dynamic_2010}%
  \BibitemOpen
  \bibfield  {author} {\bibinfo {author} {\bibfnamefont {R.}~\bibnamefont
  {Sensarma}}, \bibinfo {author} {\bibfnamefont {E.~H.}\ \bibnamefont {Hwang}},
  \ and\ \bibinfo {author} {\bibfnamefont {S.}~\bibnamefont {Das~Sarma}},\
  }\href {\doibase 10.1103/PhysRevB.82.195428} {\bibfield  {journal} {\bibinfo
  {journal} {Physical Review B}\ }\textbf {\bibinfo {volume} {82}},\ \bibinfo
  {pages} {195428} (\bibinfo {year} {2010})}\BibitemShut {NoStop}%
\bibitem [{\citenamefont {Gangadharaiah}\ \emph {et~al.}(2008)\citenamefont
  {Gangadharaiah}, \citenamefont {Farid},\ and\ \citenamefont
  {Mishchenko}}]{gangadharaiah_charge_2008}%
  \BibitemOpen
  \bibfield  {author} {\bibinfo {author} {\bibfnamefont {S.}~\bibnamefont
  {Gangadharaiah}}, \bibinfo {author} {\bibfnamefont {A.~M.}\ \bibnamefont
  {Farid}}, \ and\ \bibinfo {author} {\bibfnamefont {E.~G.}\ \bibnamefont
  {Mishchenko}},\ }\href {\doibase 10.1103/PhysRevLett.100.166802} {\bibfield
  {journal} {\bibinfo  {journal} {Physical Review Letters}\ }\textbf {\bibinfo
  {volume} {100}},\ \bibinfo {pages} {166802} (\bibinfo {year}
  {2008})}\BibitemShut {NoStop}%
\bibitem [{\citenamefont {Wunsch}\ \emph {et~al.}(2006)\citenamefont {Wunsch},
  \citenamefont {Stauber}, \citenamefont {Sols},\ and\ \citenamefont
  {Guinea}}]{wunsch_dynamical_2006}%
  \BibitemOpen
  \bibfield  {author} {\bibinfo {author} {\bibfnamefont {B.}~\bibnamefont
  {Wunsch}}, \bibinfo {author} {\bibfnamefont {T.}~\bibnamefont {Stauber}},
  \bibinfo {author} {\bibfnamefont {F.}~\bibnamefont {Sols}}, \ and\ \bibinfo
  {author} {\bibfnamefont {F.}~\bibnamefont {Guinea}},\ }\href {\doibase
  10.1088/1367-2630/8/12/318} {\bibfield  {journal} {\bibinfo  {journal} {New
  Journal of Physics}\ }\textbf {\bibinfo {volume} {8}},\ \bibinfo {pages}
  {318} (\bibinfo {year} {2006})}\BibitemShut {NoStop}%
\bibitem [{\citenamefont {Scholz}\ \emph {et~al.}(2013)\citenamefont {Scholz},
  \citenamefont {Stauber},\ and\ \citenamefont
  {Schliemann}}]{scholz_plasmons_2013}%
  \BibitemOpen
  \bibfield  {author} {\bibinfo {author} {\bibfnamefont {A.}~\bibnamefont
  {Scholz}}, \bibinfo {author} {\bibfnamefont {T.}~\bibnamefont {Stauber}}, \
  and\ \bibinfo {author} {\bibfnamefont {J.}~\bibnamefont {Schliemann}},\
  }\href {http://arxiv.org/abs/1306.1666} {\bibfield  {journal} {\bibinfo
  {journal} {{arXiv:1306.1666}}\ } (\bibinfo {year} {2013})}\BibitemShut
  {NoStop}%
\bibitem [{\citenamefont {Dresselhaus}(1955)}]{dresselhaus_spin-orbit_1955}%
  \BibitemOpen
  \bibfield  {author} {\bibinfo {author} {\bibfnamefont {G.}~\bibnamefont
  {Dresselhaus}},\ }\href {\doibase 10.1103/PhysRev.100.580} {\bibfield
  {journal} {\bibinfo  {journal} {Physical Review}\ }\textbf {\bibinfo {volume}
  {100}},\ \bibinfo {pages} {580} (\bibinfo {year} {1955})}\BibitemShut
  {NoStop}%
\bibitem [{\citenamefont {Mak}\ \emph {et~al.}(2012)\citenamefont {Mak},
  \citenamefont {He}, \citenamefont {Shan},\ and\ \citenamefont
  {Heinz}}]{mak_control_2012}%
  \BibitemOpen
  \bibfield  {author} {\bibinfo {author} {\bibfnamefont {K.~F.}\ \bibnamefont
  {Mak}}, \bibinfo {author} {\bibfnamefont {K.}~\bibnamefont {He}}, \bibinfo
  {author} {\bibfnamefont {J.}~\bibnamefont {Shan}}, \ and\ \bibinfo {author}
  {\bibfnamefont {T.~F.}\ \bibnamefont {Heinz}},\ }\href {\doibase
  10.1038/nnano.2012.96} {\bibfield  {journal} {\bibinfo  {journal} {Nature
  Nanotechnology}\ }\textbf {\bibinfo {volume} {7}},\ \bibinfo {pages} {494}
  (\bibinfo {year} {2012})}\BibitemShut {NoStop}%
\bibitem [{\citenamefont {Cao}\ \emph {et~al.}(2012)\citenamefont {Cao},
  \citenamefont {Wang}, \citenamefont {Han}, \citenamefont {Ye}, \citenamefont
  {Zhu}, \citenamefont {Shi}, \citenamefont {Niu}, \citenamefont {Tan},
  \citenamefont {Wang}, \citenamefont {Liu},\ and\ \citenamefont
  {Feng}}]{cao_valley-selective_2012}%
  \BibitemOpen
  \bibfield  {author} {\bibinfo {author} {\bibfnamefont {T.}~\bibnamefont
  {Cao}}, \bibinfo {author} {\bibfnamefont {G.}~\bibnamefont {Wang}}, \bibinfo
  {author} {\bibfnamefont {W.}~\bibnamefont {Han}}, \bibinfo {author}
  {\bibfnamefont {H.}~\bibnamefont {Ye}}, \bibinfo {author} {\bibfnamefont
  {C.}~\bibnamefont {Zhu}}, \bibinfo {author} {\bibfnamefont {J.}~\bibnamefont
  {Shi}}, \bibinfo {author} {\bibfnamefont {Q.}~\bibnamefont {Niu}}, \bibinfo
  {author} {\bibfnamefont {P.}~\bibnamefont {Tan}}, \bibinfo {author}
  {\bibfnamefont {E.}~\bibnamefont {Wang}}, \bibinfo {author} {\bibfnamefont
  {B.}~\bibnamefont {Liu}}, \ and\ \bibinfo {author} {\bibfnamefont
  {J.}~\bibnamefont {Feng}},\ }\href {\doibase 10.1038/ncomms1882} {\bibfield
  {journal} {\bibinfo  {journal} {Nature Communications}\ }\textbf {\bibinfo
  {volume} {3}},\ \bibinfo {pages} {887} (\bibinfo {year} {2012})}\BibitemShut
  {NoStop}%
\bibitem [{\citenamefont {Zeng}\ \emph {et~al.}(2012)\citenamefont {Zeng},
  \citenamefont {Dai}, \citenamefont {Yao}, \citenamefont {Xiao},\ and\
  \citenamefont {Cui}}]{zeng_valley_2012}%
  \BibitemOpen
  \bibfield  {author} {\bibinfo {author} {\bibfnamefont {H.}~\bibnamefont
  {Zeng}}, \bibinfo {author} {\bibfnamefont {J.}~\bibnamefont {Dai}}, \bibinfo
  {author} {\bibfnamefont {W.}~\bibnamefont {Yao}}, \bibinfo {author}
  {\bibfnamefont {D.}~\bibnamefont {Xiao}}, \ and\ \bibinfo {author}
  {\bibfnamefont {X.}~\bibnamefont {Cui}},\ }\href {\doibase
  10.1038/nnano.2012.95} {\bibfield  {journal} {\bibinfo  {journal} {Nature
  Nanotechnology}\ }\textbf {\bibinfo {volume} {7}},\ \bibinfo {pages} {490}
  (\bibinfo {year} {2012})}\BibitemShut {NoStop}%
\bibitem [{\citenamefont {Xiao}\ \emph {et~al.}(2012)\citenamefont {Xiao},
  \citenamefont {Liu}, \citenamefont {Feng}, \citenamefont {Xu},\ and\
  \citenamefont {Yao}}]{xiao_coupled_2012}%
  \BibitemOpen
  \bibfield  {author} {\bibinfo {author} {\bibfnamefont {D.}~\bibnamefont
  {Xiao}}, \bibinfo {author} {\bibfnamefont {G.-B.}\ \bibnamefont {Liu}},
  \bibinfo {author} {\bibfnamefont {W.}~\bibnamefont {Feng}}, \bibinfo {author}
  {\bibfnamefont {X.}~\bibnamefont {Xu}}, \ and\ \bibinfo {author}
  {\bibfnamefont {W.}~\bibnamefont {Yao}},\ }\href {\doibase
  10.1103/PhysRevLett.108.196802} {\bibfield  {journal} {\bibinfo  {journal}
  {Physical Review Letters}\ }\textbf {\bibinfo {volume} {108}},\ \bibinfo
  {pages} {196802} (\bibinfo {year} {2012})}\BibitemShut {NoStop}%
\bibitem [{\citenamefont {Ochoa}\ \emph {et~al.}(2013)\citenamefont {Ochoa},
  \citenamefont {Guinea},\ and\ \citenamefont {Fal'ko}}]{ochoa_spin_2013}%
  \BibitemOpen
  \bibfield  {author} {\bibinfo {author} {\bibfnamefont {H.}~\bibnamefont
  {Ochoa}}, \bibinfo {author} {\bibfnamefont {F.}~\bibnamefont {Guinea}}, \
  and\ \bibinfo {author} {\bibfnamefont {V.~I.}\ \bibnamefont {Fal'ko}},\
  }\href {http://arxiv.org/abs/1308.0928} {\bibfield  {journal} {\bibinfo
  {journal} {{arXiv:1308.0928}}\ } (\bibinfo {year} {2013})}\BibitemShut
  {NoStop}%
\bibitem [{\citenamefont {Hasan}\ and\ \citenamefont
  {Kane}(2010)}]{hasan_colloquium:_2010}%
  \BibitemOpen
  \bibfield  {author} {\bibinfo {author} {\bibfnamefont {M.~Z.}\ \bibnamefont
  {Hasan}}\ and\ \bibinfo {author} {\bibfnamefont {C.~L.}\ \bibnamefont
  {Kane}},\ }\href {\doibase 10.1103/RevModPhys.82.3045} {\bibfield  {journal}
  {\bibinfo  {journal} {Reviews of Modern Physics}\ }\textbf {\bibinfo {volume}
  {82}},\ \bibinfo {pages} {3045} (\bibinfo {year} {2010})}\BibitemShut
  {NoStop}%
\bibitem [{\citenamefont {Das~Sarma}\ and\ \citenamefont
  {Madhukar}(1981)}]{das_sarma_collective_1981}%
  \BibitemOpen
  \bibfield  {author} {\bibinfo {author} {\bibfnamefont {S.}~\bibnamefont
  {Das~Sarma}}\ and\ \bibinfo {author} {\bibfnamefont {A.}~\bibnamefont
  {Madhukar}},\ }\href {\doibase 10.1103/PhysRevB.23.805} {\bibfield  {journal}
  {\bibinfo  {journal} {Physical Review B}\ }\textbf {\bibinfo {volume} {23}},\
  \bibinfo {pages} {805} (\bibinfo {year} {1981})}\BibitemShut {NoStop}%
\bibitem [{\citenamefont {Santoro}\ and\ \citenamefont
  {Giuliani}(1988)}]{santoro_acoustic_1988}%
  \BibitemOpen
  \bibfield  {author} {\bibinfo {author} {\bibfnamefont {G.~E.}\ \bibnamefont
  {Santoro}}\ and\ \bibinfo {author} {\bibfnamefont {G.~F.}\ \bibnamefont
  {Giuliani}},\ }\href {\doibase 10.1103/PhysRevB.37.937} {\bibfield  {journal}
  {\bibinfo  {journal} {Physical Review B}\ }\textbf {\bibinfo {volume} {37}},\
  \bibinfo {pages} {937} (\bibinfo {year} {1988})}\BibitemShut {NoStop}%
\bibitem [{\citenamefont {Profumo}\ \emph {et~al.}(2012)\citenamefont
  {Profumo}, \citenamefont {Asgari}, \citenamefont {Polini},\ and\
  \citenamefont {{MacDonald}}}]{profumo_double-layer_2012}%
  \BibitemOpen
  \bibfield  {author} {\bibinfo {author} {\bibfnamefont {R.~E.~V.}\
  \bibnamefont {Profumo}}, \bibinfo {author} {\bibfnamefont {R.}~\bibnamefont
  {Asgari}}, \bibinfo {author} {\bibfnamefont {M.}~\bibnamefont {Polini}}, \
  and\ \bibinfo {author} {\bibfnamefont {A.~H.}\ \bibnamefont {{MacDonald}}},\
  }\href {\doibase 10.1103/PhysRevB.85.085443} {\bibfield  {journal} {\bibinfo
  {journal} {Physical Review B}\ }\textbf {\bibinfo {volume} {85}},\ \bibinfo
  {pages} {085443} (\bibinfo {year} {2012})}\BibitemShut {NoStop}%
\bibitem [{\citenamefont {Pisarra}\ \emph {et~al.}(2013)\citenamefont
  {Pisarra}, \citenamefont {Sindona}, \citenamefont {Riccardi}, \citenamefont
  {Silkin},\ and\ \citenamefont {Pitarke}}]{pisarra_acoustic_2013}%
  \BibitemOpen
  \bibfield  {author} {\bibinfo {author} {\bibfnamefont {M.}~\bibnamefont
  {Pisarra}}, \bibinfo {author} {\bibfnamefont {A.}~\bibnamefont {Sindona}},
  \bibinfo {author} {\bibfnamefont {P.}~\bibnamefont {Riccardi}}, \bibinfo
  {author} {\bibfnamefont {V.~M.}\ \bibnamefont {Silkin}}, \ and\ \bibinfo
  {author} {\bibfnamefont {J.~M.}\ \bibnamefont {Pitarke}},\ }\href
  {http://arxiv.org/abs/1306.6273} {\bibfield  {journal} {\bibinfo  {journal}
  {{arXiv:1306.6273}}\ } (\bibinfo {year} {2013})}\BibitemShut {NoStop}%
\bibitem [{\citenamefont {Cappelluti}\ \emph {et~al.}(2013)\citenamefont
  {Cappelluti}, \citenamefont {Roldan}, \citenamefont {Silva-Guillen},
  \citenamefont {Ordejon},\ and\ \citenamefont
  {Guinea}}]{cappelluti_tight-binding_2013}%
  \BibitemOpen
  \bibfield  {author} {\bibinfo {author} {\bibfnamefont {E.}~\bibnamefont
  {Cappelluti}}, \bibinfo {author} {\bibfnamefont {R.}~\bibnamefont {Roldan}},
  \bibinfo {author} {\bibfnamefont {J.~A.}\ \bibnamefont {Silva-Guillen}},
  \bibinfo {author} {\bibfnamefont {P.}~\bibnamefont {Ordejon}}, \ and\
  \bibinfo {author} {\bibfnamefont {F.}~\bibnamefont {Guinea}},\ }\href
  {http://arxiv.org/abs/1304.4831} {\bibfield  {journal} {\bibinfo  {journal}
  {{arXiv:1304.4831}}\ } (\bibinfo {year} {2013})}\BibitemShut {NoStop}%
\bibitem [{\citenamefont {Rostami}\ \emph {et~al.}(2013)\citenamefont
  {Rostami}, \citenamefont {Moghaddam},\ and\ \citenamefont
  {Asgari}}]{Rostami_Moghaddam_Asgari_2013}%
  \BibitemOpen
  \bibfield  {author} {\bibinfo {author} {\bibfnamefont {H.}~\bibnamefont
  {Rostami}}, \bibinfo {author} {\bibfnamefont {A.~G.}\ \bibnamefont
  {Moghaddam}}, \ and\ \bibinfo {author} {\bibfnamefont {R.}~\bibnamefont
  {Asgari}},\ }\href {\doibase 10.1103/PhysRevB.88.085440} {\bibfield
  {journal} {\bibinfo  {journal} {Physical Review B}\ }\textbf {\bibinfo
  {volume} {88}},\ \bibinfo {pages} {085440} (\bibinfo {year}
  {2013})}\BibitemShut {NoStop}%
\bibitem [{\citenamefont {Kormanyos}\ \emph {et~al.}(2013)\citenamefont
  {Kormanyos}, \citenamefont {Zolyomi}, \citenamefont {Drummond}, \citenamefont
  {Rakyta}, \citenamefont {Burkard},\ and\ \citenamefont
  {Falko}}]{Kormanyos_Zolyomi_Drummond_Rakyta_Burkard_Falko_2013}%
  \BibitemOpen
  \bibfield  {author} {\bibinfo {author} {\bibfnamefont {A.}~\bibnamefont
  {Kormanyos}}, \bibinfo {author} {\bibfnamefont {V.}~\bibnamefont {Zolyomi}},
  \bibinfo {author} {\bibfnamefont {N.~D.}\ \bibnamefont {Drummond}}, \bibinfo
  {author} {\bibfnamefont {P.}~\bibnamefont {Rakyta}}, \bibinfo {author}
  {\bibfnamefont {G.}~\bibnamefont {Burkard}}, \ and\ \bibinfo {author}
  {\bibfnamefont {V.~I.}\ \bibnamefont {Falko}},\ }\href {\doibase
  10.1103/PhysRevB.88.045416} {\bibfield  {journal} {\bibinfo  {journal}
  {Physical Review B}\ }\textbf {\bibinfo {volume} {88}},\ \bibinfo {pages}
  {045416} (\bibinfo {year} {2013})}\BibitemShut {NoStop}%
\bibitem [{\citenamefont {Wei}\ \emph {et~al.}(2013)\citenamefont {Wei},
  \citenamefont {Katmis}, \citenamefont {Assaf}, \citenamefont {Steinberg},
  \citenamefont {Jarillo-Herrero}, \citenamefont {Heiman},\ and\ \citenamefont
  {Moodera}}]{wei_exchange-coupling-induced_2013}%
  \BibitemOpen
  \bibfield  {author} {\bibinfo {author} {\bibfnamefont {P.}~\bibnamefont
  {Wei}}, \bibinfo {author} {\bibfnamefont {F.}~\bibnamefont {Katmis}},
  \bibinfo {author} {\bibfnamefont {B.~A.}\ \bibnamefont {Assaf}}, \bibinfo
  {author} {\bibfnamefont {H.}~\bibnamefont {Steinberg}}, \bibinfo {author}
  {\bibfnamefont {P.}~\bibnamefont {Jarillo-Herrero}}, \bibinfo {author}
  {\bibfnamefont {D.}~\bibnamefont {Heiman}}, \ and\ \bibinfo {author}
  {\bibfnamefont {J.~S.}\ \bibnamefont {Moodera}},\ }\href {\doibase
  10.1103/PhysRevLett.110.186807} {\bibfield  {journal} {\bibinfo  {journal}
  {Physical Review Letters}\ }\textbf {\bibinfo {volume} {110}},\ \bibinfo
  {pages} {186807} (\bibinfo {year} {2013})}\BibitemShut {NoStop}%
\bibitem [{\citenamefont {Yang}\ and\ \citenamefont {et.
  al.}(2013)}]{yang_2013}%
  \BibitemOpen
  \bibfield  {author} {\bibinfo {author} {\bibfnamefont {Q.~I.}\ \bibnamefont
  {Yang}}\ and\ \bibinfo {author} {\bibnamefont {et. al.}},\ }\href
  {http://arxiv.org/abs/1306.2038} {\bibfield  {journal} {\bibinfo  {journal}
  {{arXiv:1306.2038}}\ } (\bibinfo {year} {2013})}\BibitemShut {NoStop}%
\bibitem [{\citenamefont {Boschker}\ \emph {et~al.}(2012)\citenamefont
  {Boschker}, \citenamefont {Kautz}, \citenamefont {Houwman}, \citenamefont
  {Siemons}, \citenamefont {Blank}, \citenamefont {Huijben}, \citenamefont
  {Koster}, \citenamefont {Vailionis},\ and\ \citenamefont
  {Rijnders}}]{boschker_high-temperature_2012}%
  \BibitemOpen
  \bibfield  {author} {\bibinfo {author} {\bibfnamefont {H.}~\bibnamefont
  {Boschker}}, \bibinfo {author} {\bibfnamefont {J.}~\bibnamefont {Kautz}},
  \bibinfo {author} {\bibfnamefont {E.~P.}\ \bibnamefont {Houwman}}, \bibinfo
  {author} {\bibfnamefont {W.}~\bibnamefont {Siemons}}, \bibinfo {author}
  {\bibfnamefont {D.~H.~A.}\ \bibnamefont {Blank}}, \bibinfo {author}
  {\bibfnamefont {M.}~\bibnamefont {Huijben}}, \bibinfo {author} {\bibfnamefont
  {G.}~\bibnamefont {Koster}}, \bibinfo {author} {\bibfnamefont
  {A.}~\bibnamefont {Vailionis}}, \ and\ \bibinfo {author} {\bibfnamefont
  {G.}~\bibnamefont {Rijnders}},\ }\href {\doibase
  10.1103/PhysRevLett.109.157207} {\bibfield  {journal} {\bibinfo  {journal}
  {Physical Review Letters}\ }\textbf {\bibinfo {volume} {109}},\ \bibinfo
  {pages} {157207} (\bibinfo {year} {2012})}\BibitemShut {NoStop}%
\bibitem [{\citenamefont {Walther}(1967)}]{Walther_1967}%
  \BibitemOpen
  \bibfield  {author} {\bibinfo {author} {\bibfnamefont {K.}~\bibnamefont
  {Walther}},\ }\href@noop {} {\bibfield  {journal} {\bibinfo  {journal} {Solid
  State Communications}\ }\textbf {\bibinfo {volume} {5}},\ \bibinfo {pages}
  {399} (\bibinfo {year} {1967})}\BibitemShut {NoStop}%
\end{thebibliography}%

\end{document}